\documentclass[twocolumn,aps,prl,showpacs, superscriptaddress, amsmath,amssymb]{revtex4}

\usepackage{graphicx}
\usepackage{dcolumn}
\usepackage{bm}
\usepackage{epstopdf}
\usepackage{color}
\usepackage{pdfpages}

\begin{document}

\title{Dynamic self-assembly of charged colloidal strings and walls in simple fluid flows}
\author{Yu Abe}
\affiliation{Department of Chemical Engineering and Materials
Science, University of Minnesota, Minneapolis, MN 55455}
\affiliation{Films $\&$ Film Products Research Laboratories, Toray Industries, Inc.
1-1, Sonoyama 1-chome, Otsu, Shiga, 520-8558, Japan}
\author{Bo Zhang}
\affiliation{Department of Chemical Engineering and Materials
Science, University of Minnesota, Minneapolis, MN 55455}
\author{Leonardo Gordillo}
\affiliation{Department of Chemical Engineering and Materials
Science, University of Minnesota, Minneapolis, MN 55455}
\author{Alireza Mohammad Karim}
\affiliation{Department of Chemical Engineering and Materials
Science, University of Minnesota, Minneapolis, MN 55455}
\author{Lorraine F. Francis}
\affiliation{Department of Chemical Engineering and Materials
Science, University of Minnesota, Minneapolis, MN 55455}
\author{Xiang Cheng}
\affiliation{Department of Chemical Engineering and Materials
Science, University of Minnesota, Minneapolis, MN 55455}
\date{\today}

\begin{abstract}

Colloidal particles can self-assemble into various ordered structures in fluid flows that have potential applications in biomedicine, materials synthesis and encryption. These dynamic processes are also of fundamental interest for probing the general principles of self-assembly in non-equilibrium conditions. Here, we report a simple microfluidic experiment, where charged colloidal particles self-assemble into flow-aligned 1D strings with regular particle spacing near a solid boundary. Using high-speed confocal microscopy, we systematically investigate the influence of flow rates, electrostatics and particle polydispersity on the observed string structures. By studying the detailed dynamics of stable flow-driven particle pairs, we quantitatively characterize interparticle interactions. Based on the results, we construct a simple model that explains the intriguing non-equilibrium self-assembly process. Our study shows that the colloidal strings arise from a delicate balance between attractive hydrodynamic coupling and repulsive electrostatic interaction between particles. Finally, we demonstrate that, with the assistance of transverse electric fields, a similar mechanism also leads to the formation of 2D colloidal walls.              

\end{abstract}

\maketitle

\section{I. Introduction}

Self-assembly of colloidal particles is a process of growing interest because of its broad application in fields ranging from photonics to materials synthesis and to biomedical engineering \cite{Lee10,Cong13}. Although the mechanisms of colloidal self-assembly in equilibrium have been well explored with many interesting colloidal structures discovered \cite{Clark79,Leunissen05,Zerrouki08,Chen11}, our understanding of non-equilibrium self-assembly is still undeveloped. Out-of-equilibrium dynamic self-assembly provides not only a powerful tool for designing novel colloidal materials with unusual mesoscopic structures, but also a fertile ground for exploring non-equilibrium statistical mechanics. Among all different non-equilibrium systems, flow-driven particle self-assembly attracts probably the most attention due to the rapid growth of microfluidic techniques \cite{Vermant05,Haw98,Cohen04,Wu09,Besseling12}, where interactions between fluid flows and colloidal particles are frequently encountered.   

Particularly, the flow-driven self-assembly of one dimensional (1D) colloidal strings with regular particle spacing shows great potential in microfluidic applications \cite{Montesi04,Osuji08,Lee10_1,Cheng12,Pasquino13,Lin14}. 1D colloidal strings offer a precise control over the positions of colloidal particles such as cells and droplets, which is crucial for flow cytometry, cell separation and diagnostics \cite{Huh05,Tewhey09,Patra16}, and microfluidic computation and encryption \cite{Fuerstman07,Prakash07,Hashimoto09}. Nevertheless, compared with a myriad of methods for creating 2D or 3D colloid structures, fewer approaches are available for assembling 1D colloidal strings. Due to the isotropic nature of interparticle interactions, it is difficult to grow colloidal spheres along a single direction. Thus, 1D colloidal strings are usually constructed through special processes. In equilibrium, colloidal strings have been grown via template-directed colloidal epitaxy \cite{Malaquin07} and electromagnetic field-assistant self-assembly \cite{Zerrouki08,Vutukuri12,Wang13}. In fluid flows, the self-assembly of colloidal strings aligned along the vorticity direction of shear flows has been reported when the colloidal particles are subjected to shear in viscoelastic fluids \cite{Montesi04,Osuji08,Pasquino13} and in Newtonian fluids under strong confinement \cite{Cheng12,Lin14}. Flow-aligned colloidal strings that are more useful in microfluidic applications have been observed at finite Reynolds numbers (Re $> 1$) when the effect of inertia is significant.\cite{Lee10_1} Unfortunately, the high-Re condition cannot be easily satisfied in many microfluidic devices. As a result, instead of relying on the process of self-assembly, flow-aligned colloidal strings are normally created manually in microfluidics, where particles are injected into microfluidic channels at externally-controlled regular time intervals \cite{Beatus06,Dendukuri09,Shen16}. The string structures thus formed are metastable, susceptible to external perturbations \cite{Beatus06}.    

In this paper, we report a simple experimental method for constructing 1D colloidal crystals, where, under unidirectional flows, charged colloidal particles self-assemble into flow-aligned strings with regular particle spacing near a solid boundary. Using fast confocal microscopy, we study the dynamics of the self-assembly process. By systematically varying key control parameters such as the flow rate, the strength of electrostatic interactions, the size and polydispersity of particles and the dimension of microfluidic channels, we delineate the condition for the formation of colloidal strings. Based on our experimental observation, we propose a simple model that explains the intriguing non-equilibrium self-assembly process. Although our model is minimalistic with crude approximations, it is capable of capturing the essential mechanism for the origin of colloidal strings. As such, our experiments demonstrate the possibility of assembling flow-aligned 1D ordered colloidal structures at low Re through the balance of hydrodynamic coupling and electrostatic repulsion. Compared with previous methods, our study provides a much simpler approach for constructing a large number of stable flow-aligned colloidal strings, potentially useful for different microfluidic applications. Lastly, we show that this approach also leads to the formation of 2D colloidal walls when a transverse electric field is applied in the same colloidal system. 

\section{II. Experiments}
We synthesized poly(methyl methacrylate) (PMMA) spheres as our colloidal particles \cite{Antl86,Pathmamanoharan97}, which were fluorescently labeled with Nile red. The particles were stabilized by covalently bound polymer brushes made of poly(1,2-hydroxy stearic acid) (PHSA). Most our experiments were conducted using particles of average diameter $d = 1.36$ $\mu$m, although two other batches of particles with $d = 1.56$ $\mu$m and 1.86 $\mu$m were also used to test the effect of polydispersity. The polydispersities of particles in these particle batches are below $11\%$. Particle sizes and polydispersities were determined from dynamic light scattering (DLS). PMMA particles were suspended in a mixture of decahydronaphthalene (decalin, 27.2 wt$\%$) and cis + trans cyclohexyl bromide (CXB, 72.8 wt$\%$), which matches both the density and refractive index of the particles. The density and the viscosity of the mixture are $\rho = 1.20$ g/cm$^3$ and $\eta = 1.74$ mPa$\cdot$s, respectively. Its relative dielectric constant is $\epsilon_r = 5.6$. The volume fraction of suspensions was fixed at $\phi = 1\%$ for most of our experiments. We obtained the surface charge of PMMA particles in the decalin/CXB mixture via electrophoresis \cite{Linden15}. The electrophoretic mobility of particles in a DC field of 6 V/mm was measured, which can be converted into the zeta potential and the surface charges of particles based on a theory by Carrique {\it et al.} \cite{Carrique02}. The numbers of positive surface charges on the three batches of PMMA particles with increasing $d$ are similar with $Z = 158 \pm 8.2$, $152 \pm 12$ and $160 \pm 4.7$, respectively. Note that we used a Debye screening length $\kappa^{-1} \approx 1.3$ $\mu$m in our calculation, which is derived from the measured conductivity of the mixture at 218 pS/cm \cite{Linden15}. The screening length can be changed when salt is added into the solvent. All our measurements were conducted at room temperature.

\begin{figure}
\begin{center}
\includegraphics[width=3.2in]{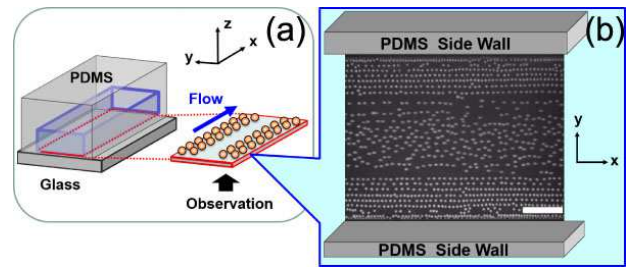}
\end{center}
\caption[Examples of strings]{Formation of colloidal strings in microfluidic flows. (a) A schematic showing the geometry of our experiments. A coordinate system is defined in the upper left. (b) An example of colloidal strings in the narrow channel next to the glass bottom (see also Supplementary Video 1). Flow is along the $x$ direction. The average diameter of particles is 1.36 $\mu$m and the flow rate is 3.0 $\mu$L/min. The scale bar is 25 $\mu$m.} \label{figure1}
\end{figure}      

\begin{figure*}
\begin{center}
\includegraphics[width=6.0in]{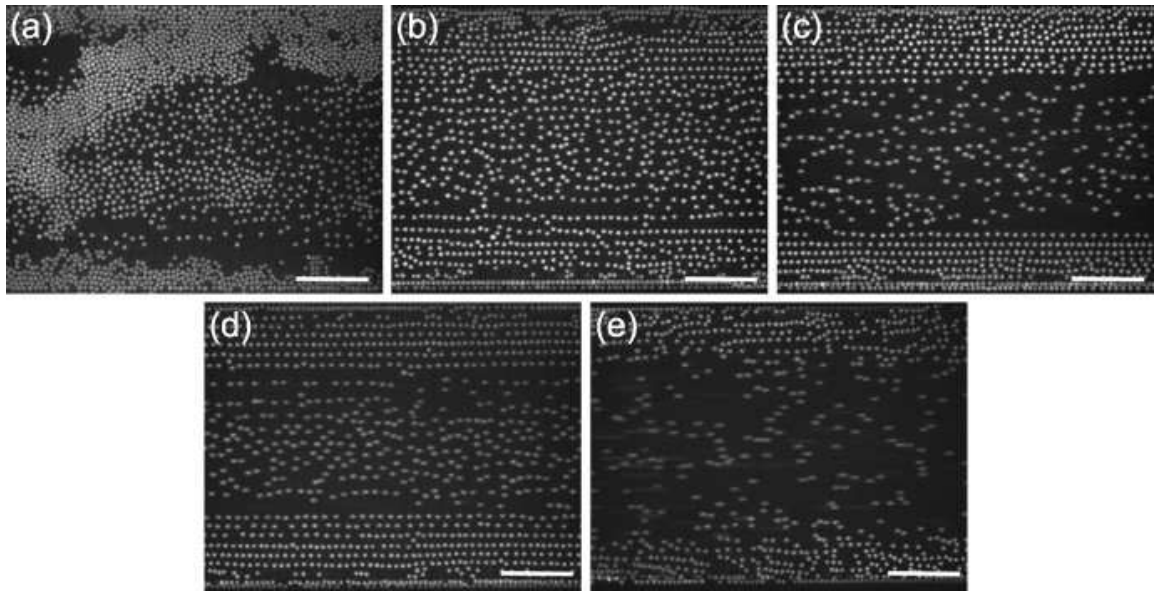}
\end{center}
\caption[Flow rate dependence]{Colloidal strings at different flow rates in the narrow channel. Suspensions flow toward the right. From (a) to (e), the flow rates are 2.0, 2.2, 2.6, 3.0 and 3.6 $\mu$L/min, respectively. The top and the bottom edges of the images correspond to the two side walls of the channel. The scale bars are 25 $\mu$m.} \label{figure2}
\end{figure*}

We fabricated microfluidic channels from poly(dimethylsiloxane) (PDMS) using conventional photolithography. The channels have rectangular cross-sections with two different inner dimensions. A narrow channel has a width of $W = 99.6 \pm 1.4$ $\mu$m, whereas a wide channel has a width of $W = 299.7 \pm 2.1$ $\mu$m. All channels have a similar height of $H = 104.8 \pm 9.2$ $\mu$m and a length of $L \sim 4$ cm. We defined a Cartesian coordinate system such that $x$ is the flow direction along the channel length, $y$ is along the width direction with $-W/2 \leq y \leq W/2$ and $z$ is along the height direction with $0 \leq z \leq H$ (Fig.~\ref{figure1}a). The cell was sealed at the bottom by a glass microscope coverslip, allowing for direct confocal imaging. Colloidal suspensions were injected into the channels using a syringe pump with a controlled volumetric flow rate ranging from 2.0 to 9.0 $\mu$L/min. The uncertainties on flow rates are less than $0.5\%$. Within this range, the particle Reynolds number, Re $\equiv \rho v_p d/\eta < 5 \times 10^{-4}$, where $v_p$ is the velocity of particles near the bottom wall. Finally, to probe the effect of transverse electric fields on the self-assembly process, we also fabricated a small microfluidic channel of 100 $\times$ 15 $\mu$m$^2$ (W $\times$ H), where the bottom and top walls of the channel were made of indium tin oxide (ITO) coated glasses. The details of the ITO channel can be found in Sec. III.E.  

We used an inverted spinning-disk confocal microscope to image the dynamics of particles near the center of a microfluidic channel away from its inlet and outlet. The images were recorded at 100 frames per second. Particle dynamics at different heights $z$ above the bottom wall were taken by varying the focal plane of a 60$\times$ microscope lens (NA 1.4). To follow the dynamics of colloidal particles for the longest possible time, we translated the sample holder of the microscope in the direction opposite to the flow of suspensions at a controlled speed in certain experiments, so that particles at a given position can be imaged in their co-flowing frame. The positions of all particles in the field of view were then tracked over a long time using a widely-used particle tracking algorithm \cite{Crocker96}.

\section{III. Results and discussions}

A typical example of colloidal strings formed in our microfluidic channel is shown in Fig. \ref{figure1}b, where PMMA particles self-assemble into flow-aligned 1D strings at the boundary next to the glass bottom (see also Supplementary Video 1). The strings move along with the ambient flow but with significantly reduced speeds. Above the bottom layer, strings disappear, where particles simply move passively with the ambient flow as expected (Fig. S1a in Supporting Information (SI)). The strings were also observed next to the top wall (Fig. S1b in SI). In this study, we focus on the dynamics of particles next to the glass bottom at $z \approx d/2$ for the convenience of imaging. Nevertheless, the mechanism discussed should equally apply for strings next to the top wall. To reveal the origin of colloidal strings, we first investigate the influence of flow rates and the strength of electrostatic interactions on the string structure. 

\subsection{A. Flow rate and ionic strength dependence}

Flow rates show a strong effect on the formation of colloidal strings (Fig.~\ref{figure2}). In the narrow channel, at a low flow rate of 2.0 $\mu$L/min, most PMMA particles entering the channel near the glass bottom stick to the glass bottom resisting the drag from the ambient flow. Since the refractive index of the particles is close to that of the glass bottom, the effective Hamaker constant and, therefore, the van der Waals attraction between particles and the glass surface should be small \cite{Russel89}. The adhesion mainly arises from the electrostatic attraction as we shall demonstrate below. At this low flow rate, particles are attracted to the glass surface and form immobile clusters instead of moving strings (Fig.~\ref{figure2}a). At a slightly higher flow rate of 2.2 $\mu$L/min, strings start to form on the surface of the glass (Fig.~\ref{figure2}b). At even higher flow rates of 2.6 $\mu$L/min and 3.0 $\mu$L/min, the string structure can be clearly identified near the side walls (Figs.~\ref{figure2}c and d). Short and less regular strings can also be observed near the center of the channel. The spacings between the strings along the $y$ direction show large variations. The average spacing between the strings decreases with increasing particle concentrations (Fig. S2 in SI). It is worth noting that the strings are very stable and robust against strong perturbations such as the presence of stuck particles on the surface and the variation of fluid flows (see Supplementary Videos 1, 2, 4 and 5). However, as we increase the flow rate further to 3.6 $\mu$L/min, colloidal strings collapse (Fig.~\ref{figure2}e). At this high flow rate, particles move passively with the ambient flow.

\begin{figure}
\begin{center}
\includegraphics[width=3in]{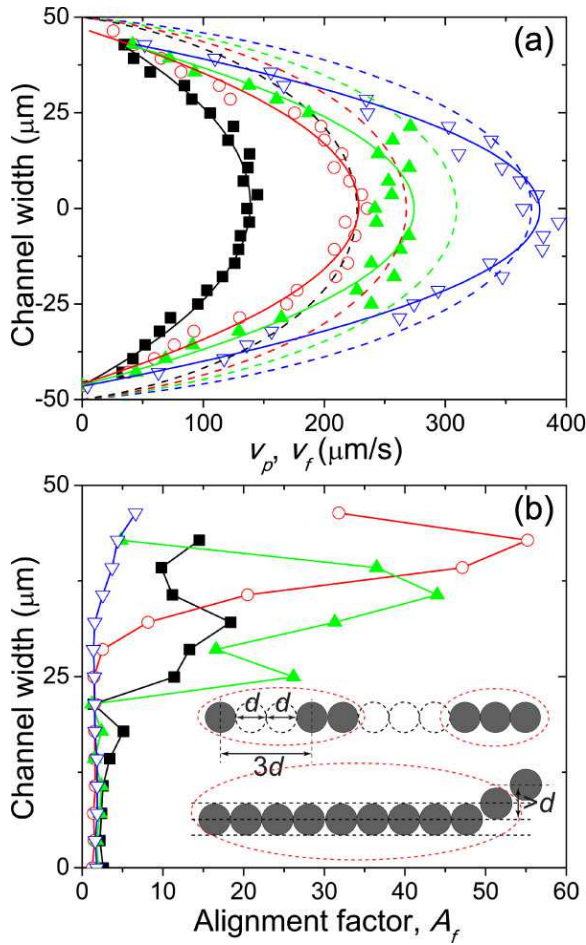}
\end{center}
\caption[Examples of strings]{Degree of the string order at different flow rates. (a) Velocity of particles, $v_p$, across the width of the channel. From left to right, the flow rates are 2.2 (black squares), 2.6 (red circles), 3.0 (green up-pointing triangles) and 3.6 $\mu$L/min (blue down-pointing triangles), respectively. The solid lines are fits with parabolic functions. The dashed lines are the calculated velocity of the unperturbed flows at $z = d/2$. (b) Alignment factor, $A_f$, across the width of the channel. Since the string order is symmetric, the results are shown only for the upper half of the channel. The symbols are the same as those used in (a). The inset illustrates the definition of strings. Particles in a dashed circle belong to one string. Neighboring particles in a string should have distances smaller than $3d$ along the $x$ direction and smaller than $d$ along the $y$ direction.} \label{figure3}
\end{figure}

To quantify the degree of the 1D string order across the width of the channel, we measure the alignment factor at different $y$, $A_f(y)$, defined as \cite{Pasquino13}
\begin{equation}
A_f = \frac{\sum_{L=1}^{\infty} N_L L^2}{\sum_{L=1}^{\infty} N_L L},
\label{equation1}
\end{equation}  
where $N_L$ is the number of strings that contain $L$ particles.  $A_f$ is similar to the weight-average molecular weight of polymer chains. $A_f \geq 1$ and $A_f = 1$ when there are only isolated particles. Note that a string is defined such that all the neighboring particles in the string are within three particle diameters away from each other along the $x$ direction and within one particle diameter away along the $y$ direction (Fig.~\ref{figure3}b lower inset). 

\begin{figure}
\begin{center}
\includegraphics[width=3 in]{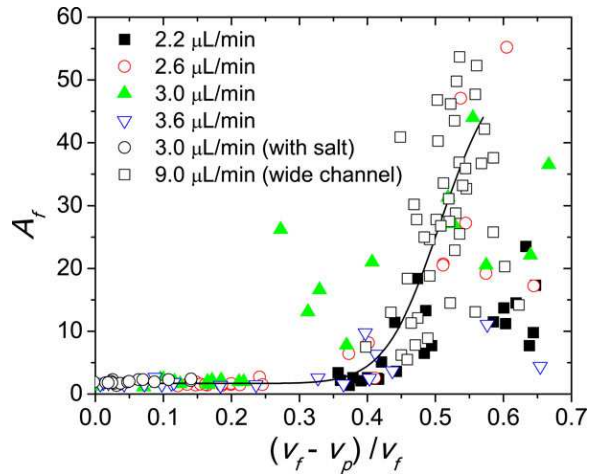}
\end{center}
\caption[Alignment factor]{Alignment factor, $A_f$, as a function of the normalized velocity difference $(v_f - v_p)/v_f$. $v_p$ is particle velocity. $v_f$ is the velocity of unperturbed ambient flows. Flow rates are indicated in the plot. The solid line provides a visual guide.} \label{figure4}
\end{figure}

The alignment factor, $A_f(y)$, the velocity of unperturbed ambient flows, $v_f(y)$, and the velocity of particles, $v_p(y)$, at different flow rates are shown in Fig.~\ref{figure3}. Here, $v_f$ is calculated based on the flow profiles of pressure-driven flows in rectangular ducts of given cross-sections at given flow rates \cite{White91}. No-slip boundary conditions at the walls of the microfluidic channels are assumed in these calculations. We verify our calculation by directly comparing the calculated velocity profiles with the measured velocity profiles above the glass bottom (Fig. S5 in SI). The relative error on $v_f$ is about $7\%$ (SI).  The velocity of particles, $v_p$, is measured from experiments via particle tracking velocimetry. Note that $v_p$ reduces to zero at $y$ slightly different from $-W/2$ and $W/2$ (Fig.~\ref{figure3}a). As shown in Fig.~\ref{figure2}, a layer of particles stick to the side walls of the channel, leading to zero particle velocity near the side walls next to the glass bottom. However, away from the glass bottom, fewer particles stick to the side walls (Fig. S1a in SI).

As the flow rate increases, the velocity difference between the particles and the unperturbed flow, $(v_f - v_p)$, decreases (Fig.~\ref{figure3}a). The degree of the string order is strongest near the side walls of the channel, which increases with the flow rate at low flow rates and decreases at high flow rates (Fig.~\ref{figure3}b). More importantly, we find a positive correlation between the degree of the string order and the velocity difference between the particles and the ambient flows (Fig.~\ref{figure4}), when the velocity difference is small at $(v_f - v_p)/v_f < 0.55$. The strings disappear when $v_f \approx v_p$. At larger velocity differences, the degree of the string order may decrease, although the data show strong fluctuations in this limit.  

\begin{figure}
\begin{center}
\includegraphics[width=1.8in]{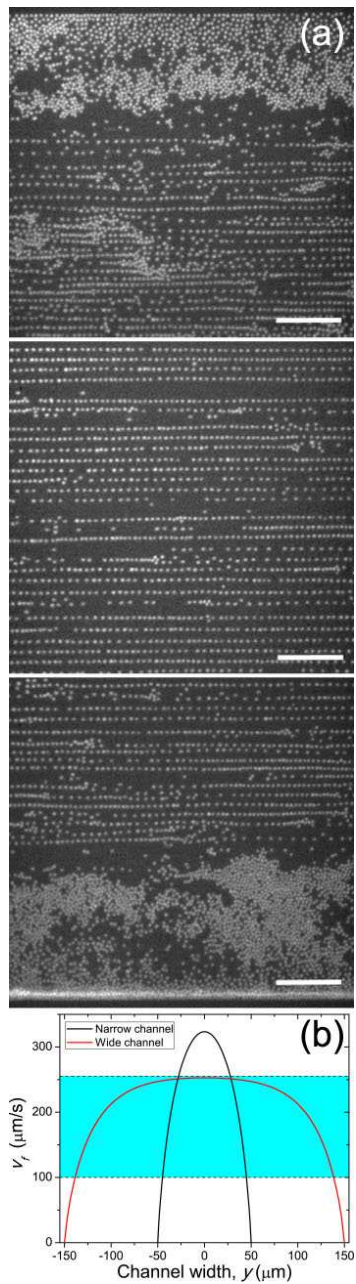}
\end{center}
\caption[Examples of strings]{(a) Formation of colloidal strings in the wide channel. The suspension flows toward the right and the flow rate is 9.0 $\mu$L/min. Since the width of the channel is larger than the field of view of our 60$\times$ lens, the entire channel is imaged by tiling three separate images. The top and bottom edges of the image correspond to the two side walls of the channel. The scale bars are 25 $\mu$m. (b) Flow profiles in the narrow and wide channels at $z = d/2$. The flow rates are 3 $\mu$L/min in the narrow channel and 9 $\mu$L/min in the wide channel, which give the same flow rate per unit cross-section area. The blue area indicates qualitatively the region where the ambient flow velocities are in the right range for the formation of colloidal strings.} \label{figure5}
\end{figure}

It is worth noting that the presence of the side walls and the shear gradients near the side walls are not necessary for the formation of the strings. We have conducted experiments in the wide channel of 300 $\mu$m in width (Sec. 2). The formation of strings can be clearly observed at the center of the channel (Fig.~\ref{figure5} and Supplementary Video 2). The difference in particle structures in the narrow and wide channels can be explained from different velocity profiles in the two channels. With a fixed flow rate per unit cross-section area, the flow velocity near the center of the narrow channel is larger than that near the center of the wide channel. We show the velocity profiles in the two channels at a flow rate of $3 \times 10^{-4}$ $\mu$L/(min$\cdot\mu$m$^2$) (Fig.~\ref{figure5}b), which corresponds to 3 $\mu$L/min in the narrow channel (the flow rate used in Fig.~\ref{figure2}d) and 9 $\mu$L/min in the wide channel (the flow rate used in Fig.~\ref{figure5}a). The velocity difference between particles and the ambient flow necessary for the formation of colloidal strings only establishes at intermediate ambient velocities, as illustrated qualitatively by the blue region in Fig.~\ref{figure5}b. At lower ambient velocities below the blue region, the drag forces on particles are too weak. Particles stick to the glass bottom (Fig.~\ref{figure2}a). At higher ambient velocities above the blue region, particles move with the ambient flow and the velocity difference reduces to zero (Fig.~\ref{figure2}e). Based on the above picture, the strings form near the wall in the narrow channel (Fig.~\ref{figure5}b). There exist small regions of one or two particle layers next to both side walls, where particles are immobile due to weak ambient flows (Fig.~\ref{figure3}a). Near the center of the narrow channel, the flow velocities are too large and the velocity differences between particles and the flow are too small to form colloidal strings (Fig.~\ref{figure2}d). In contrast, in the wide channel, the flow velocities away from the side walls all fall in the blue region. As such, colloidal strings form near the center of the wide channel. Particles are immobile in a larger region near the side walls due to the smaller velocity gradients near the walls of the wide channel (Fig.~\ref{figure5}b). Quantitatively, the correlation between $A_f$ and $(v_f-v_p)/v_f$ in the wide channel shows the similar trend as that in the narrow channel (Fig.~\ref{figure4}). Thus, our results indicate that the velocity difference between the particles and the ambient flows, instead of the shear gradient, is the key factor dictating the formation of the strings.        

\begin{figure}
\begin{center}
\includegraphics[width=3.3in]{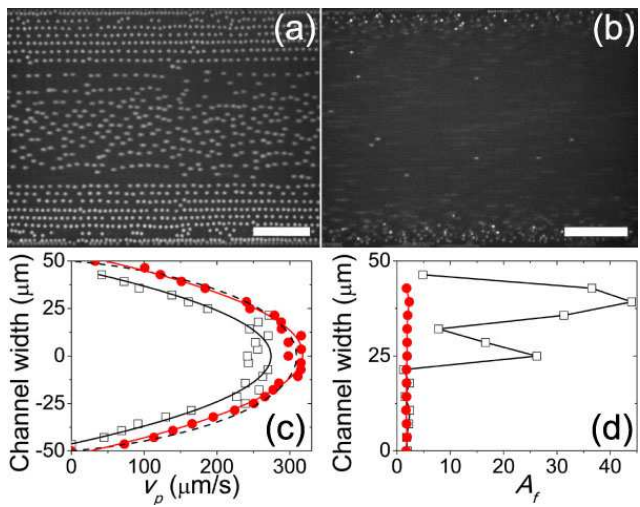}
\end{center}
\caption[Dependence of ionic strength]{Colloidal structures at different ionic strength. (a) Colloidal strings in the solvent without TBAB. (b) The collapse of strings in the solvent with 100 $\mu$M TBAB. The flow rate is 3.0 $\mu$L/min. A comparison of particle velocity and alignment factor in the two cases are shown in (c) and (d), respectively. Empty squares are for the solvent without TBAB. Red disks are for the solvent with TBAB. The scale bars are 25 $\mu$m.} \label{figure6}
\end{figure}

The formation of strings also depends on the electrostatics of particles. To reveal the effect of electrostatic interactions, we added 100 $\mu$M quaternary ammonium salt---tetrabutyl ammonium bromide (TBAB)---into the solvent, which according to a previous study reverses the sign of the surface charge of particles \cite{Royall03}. We also directly confirm the reverse of the surface charge on PMMA particles in our experiments using electrophoresis. As a result, when the salt is added, PMMA particles do not stick to the glass bottom even at low flow rates. Due to the repulsive, instead of attractive, interactions between particles and wall, the density of particles at the bottom is dramatically reduced (Fig.~\ref{figure6}b). We cannot observe the formation of strings at any flow rates (Figs.~\ref{figure6}b and d). Particle velocity also significantly increases. At 3.0 $\mu$L/min, the velocity of particles is almost the same as that of the unperturbed ambient flow (Fig.~\ref{figure6}c). The finding indicates that without salt the electrostatic attraction between particles and the glass bottom induces frictional or lubrication forces on the particles, which slow down the motion of the particles relative to the ambient flow. With the reversed electrostatic interaction at high ionic strength, the repulsion between particles and the glass bottom eliminates the drag force from the bottom. As a result, the particles move passively with the flow with $(v_f-v_p) \approx 0$. The strings do not form (Fig.~\ref{figure6}).

In summary, our experiments show that the velocity differences between the ambient flows and particles are important for the formation of colloidal strings. Such velocity differences lead to hydrodynamic coupling between particles, crucial for the bonding of particles in the strings. Moreover, the result with reversed electrostatic interactions suggests that the retarded motion of particles is induced by the electrostatic attraction between particles and the glass bottom. Charges on particles necessarily gives rise to electrostatic repulsions between particles, a factor also needs to be considered in modeling colloidal strings.    

\subsection{B. Dynamics of stable particle pairs}

\begin{figure}
\begin{center}
\includegraphics[width=3 in]{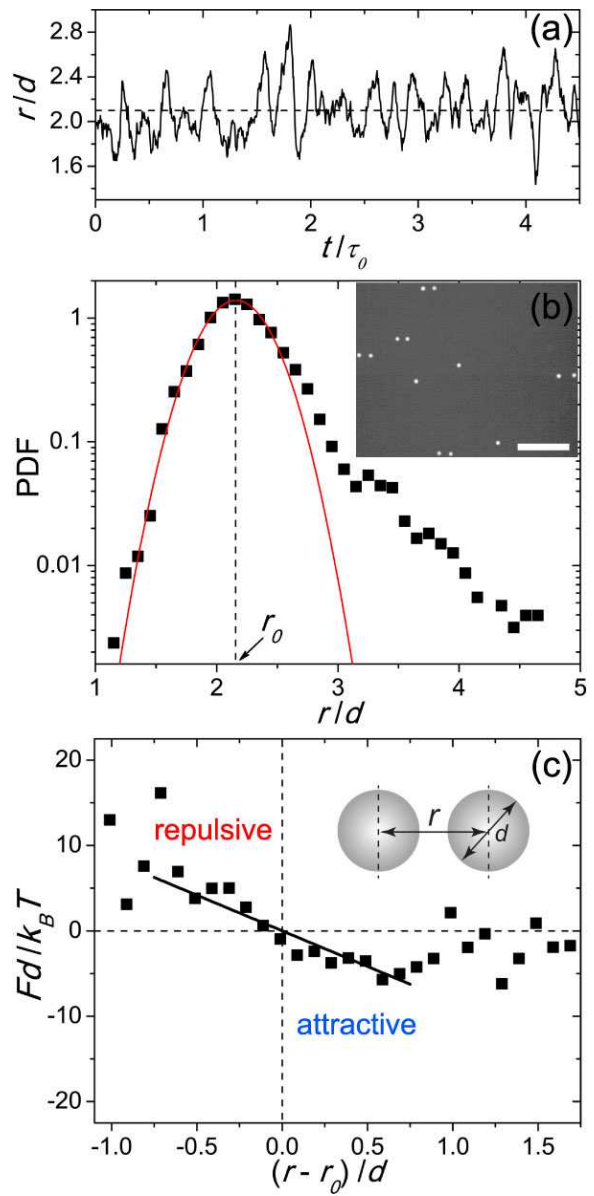}
\end{center}
\caption[Dynamics of particle pairs]{Dynamics of particle pairs in the wide channel. (a) Center-to-center distance between two particles in particle pairs, $r$, as a function of time, $t$. $r$ is normalized by the average particle diameter $d$. $r$ and $d$ are defined in the inset of (c). $t$ is normalized by the Brownian relaxation time of particles in the dilute limit, $\tau_0 = 6\pi\eta a^3/k_BT = 2.19$ s, where $\eta = 1.74$ mPa$\cdot$s is the viscosity of the solvent and $a=d/2$ is the average radius of particles. The horizontal dashed line indicates the mean distance between particles. The flow rate is 2.6 $\mu$L/min. (b) Probability distribution function (PDF) of $r$. The solid line is a fit of a Gaussian distribution using the data near the peak $r_0$ and $r<r_0$. The vertical dashed line indicates the peak of the distribution, $r_0$. The inset shows stable particle pairs in our experiments. The scale bar is 25 $\mu$m. (c) Effective force between particles, $F(r)$. $F$ is normalized by $k_BT/d$. The black line shows the model result (see text for details).} \label{figure7}
\end{figure}

To quantify the interparticle interactions, we study the dynamics of stable particle pairs. The formation of particle pairs is the precursor and the first step for the assembly of longer colloidal strings. To create stable particle pairs without interference from longer colloidal strings, we reduce the volume fractions of suspensions to $\phi = 0.01\%$ in our experiments. Most particles form stable particle pairs in the wide microfluidic channel at such a low particle concentration (Fig.~\ref{figure7}b inset). The dynamics of stable particle pairs next to the glass bottom around $y=0$ in the wide channel are then measured, where the variation of the speeds of the ambient flow is small thanks to the small velocity gradients near $y=0$. We track the center-to-center distance between two particles in particle pairs, $r$, as a function of time (Figs.~\ref{figure7}a and c inset).  Figure~\ref{figure7}b shows the probability distribution function (PDF) of the distance. The maximal probability occurs at $r_0/d = 2.160 \pm 0.006$. Near $r_0$ and at short separations with $r<r_0$, the distribution can be well approximated by a Gaussian distribution with a standard deviation of $\sigma/d = 0.52 \pm 0.01$. The PDF shows a long tail at large $r$, indicating higher probabilities to find particle pairs at large separations. 

\begin{figure}
\begin{center}
\includegraphics[width=3 in]{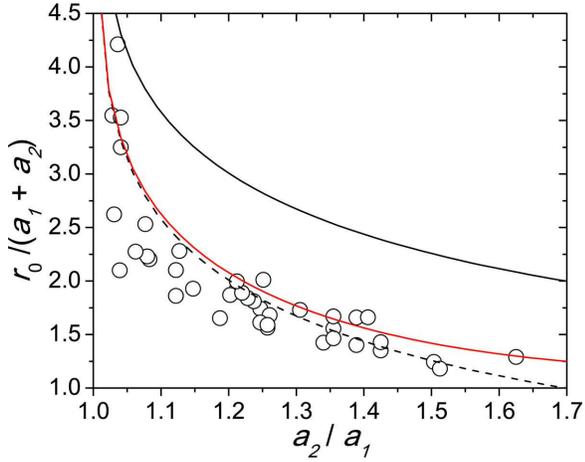}
\end{center}
\caption[Size ratio dependence]{Average center-to-center distance, $r_0$, between two particles in particle pairs of different size ratios, $a_2/a_1$. $r_0$ is normalized by the mean diameter of particle pairs, $a_1+a_2$. The black solid line is the normalized $r_0$ from the model. The dashed line shows the normalized $\lbrack r_0-(a_1+a_2) \rbrack$ from the model. The red solid line shows the normalized $r_0$ from the model when the effective distance $r_e = \lbrack r_0-(a_1+a_2) \rbrack$ is used in Eq.~\ref{equation6}.} \label{figure8}
\end{figure}

\begin{figure}
\begin{center}
\includegraphics[width=3 in]{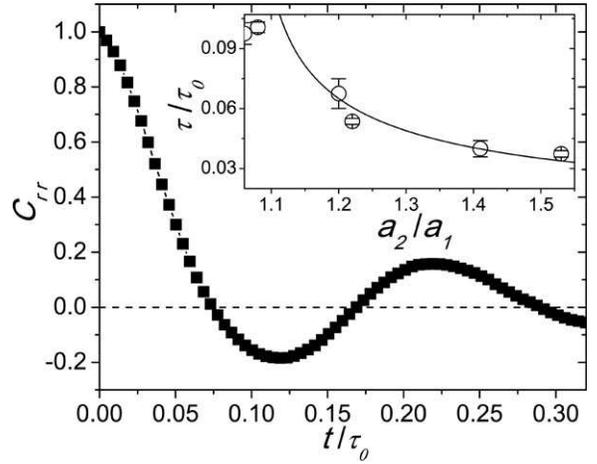}
\end{center}
\caption[Autocorrelation]{Autocorrelation of the center-to-center distance, $r(t)$, of a particle pair. The flow rate is 3.0 $\mu$L/min. The size ratio of the two particles is $a_2/a_1 = 1.2$. Inset shows the correlation time, $\tau$, as a function of $a_2/a_1$. $\tau$ is normalized by the Brownian relaxation time of particles, $\tau_0$. The solid line is a fit from the model.} \label{figure9}
\end{figure}

In analogy to equilibrium systems, the Gaussian distribution near $r_0$ allows us to define an effective potential and, accordingly, an effective force between particles in this non-equilibrium flow-driven system.       
\begin{eqnarray}
&& \frac{U(r)}{k_BT} = \ln A - \ln(\text{PDF}(r)), \nonumber \\
&& F(r) = -\frac{dU(r)}{dr} = \frac{k_BT}{\text{PDF}}\frac{d\text{PDF}}{dr},
\label{equation2}
\end{eqnarray}  
where $U(r)$ is the effective potential in the unit of thermal energy $k_BT$ and $F(r)$ is the effective force between particles in a particle pair. $A$ is the normalization constant of the probability distribution.  $F(r)$ in the unit of $k_BT/d$ is shown in Fig.~\ref{figure7}c. Near the equilibrium position $r=r_0$ where $F(r=r_0)=0$, the harmonic potential gives rises to a spring-like force. When $r<r_0$, a repulsive force pushes the two particles apart, whereas at $r>r_0$ an attraction pulls the particles together. The force can be as large as several $k_BT/d$ and, therefore, can effectively suppress the thermal motion of particles and maintain the stability of the particle pair. When the distance between particles is one particle diameter away from the equilibrium position, $r > r_0 + d$, $F(r) \approx 0$. The particles are decoupled.

The results shown in Figs.~\ref{figure7}b and c are obtained by averaging tens of particle pairs of different size ratios. A more detailed analysis shows that the equilibrium position $r_0$ depends on the size ratio of the two particles in a particle pair, $a_2/a_1$, where $a_1$ and $a_2$ are the radii of the leading and the trailing particles in the particle pair, respectively. Note that when calculating $a_2/a_1$, a correction has been made to consider the effect that the centers of small and large particles sit at slightly different heights above the glass bottom (Fig. S4 in SI). We estimate the errors on $a_2/a_1$ to be about $4\%$. To probe a larger range of size ratios, we mix the $d=1.36$ $\mu$m particle batch with two other batches of particles with $d=1.56$ and $1.86$ $\mu$m (Sec. II). The polydispersity of these particle batches allows us to obtain an almost continuous range of size ratios. We show $r_0$ as a function of $a_2/a_1$ in Fig.~\ref{figure8}. $r_0$ decreases with increasing $a_2/a_1$. The change is most drastic at small $a_2/a_1$ when the size difference of the two particles in a particle pair is small. Note that when we set the average size ratio of particle pairs equal to the polydispersity of the particle batch, i.e., $a_2/a_1 = 1.11$, we have $r_0 \approx 2.2$, consistent with our average measurements shown in Fig.~\ref{figure7}b.

We also calculated the autocorrelation of the particle-pair dynamics (Fig.~\ref{figure9}),
\begin{equation}
C_{rr}(t) = \frac{\langle\left(r(t+t_0)-\langle r \rangle\right)\left(r(t_0)-\langle r \rangle\right)\rangle}{\langle r^2 \rangle - \langle r \rangle^2}.
\label{equation3}
\end{equation}   
The average $\langle ... \rangle$ is taken over all the initial time $t_0$. $\langle r \rangle \approx r_0$. The vibration of the two particles in a particle pair around $r_0$ leads to the oscillation of $C_{rr}$ at large $t$. The correlation time of the particle-pair dynamics, $\tau$, can be defined by fitting $C_{rr}$ at small $t$ with an exponential function, $\exp(-t/\tau)$. $\tau$ indicates the rigidity of the harmonic potential in Eq.~\ref{equation2}. A smaller $\tau$ corresponds to a narrower and stiffer potential well. We find that $\tau$ is also a function of $a_2/a_1$ (Fig.~\ref{figure9} inset). A larger $a_2/a_1$ gives rise to a smaller $\tau$ and, therefore, a stiffer potential.   

\begin{figure}
\begin{center}
\includegraphics[width=3 in]{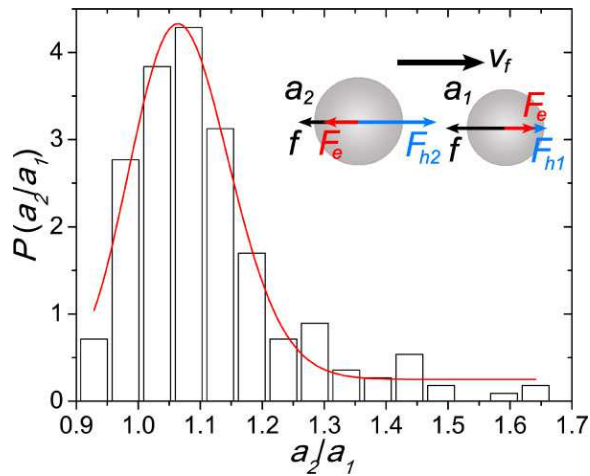}
\end{center}
\caption[Size ratio]{Probability distribution of the size ratio between the trailing and leading particles, $P(a_2/a_1)$. The red line is a fit using a log-normal function with the peak value at $a_2/a_1 = 1.064$. Inset shows the force diagram on a particle pair. The flow direction is indicated by $v_f$.} \label{figure10}
\end{figure}

The dependence of $r_0$ and $\tau$ on $a_2/a_1$ indicates that particle polydispersity plays an important role in the formation of stable particle pairs. Indeed, we measured the size ratio between the leading particle and the trailing particle, $a_2/a_1$, over 220 particle pairs. Figure~\ref{figure10} shows the probability distribution function of $a_2/a_1$. The distribution is clearly biased toward $a_2/a_1 > 1$ with the maximal probability occurring at $a_2/a_1 = 1.064$. In other words, the leading particle is smaller than the trailing particle in a particle pair.      

\subsection{C. A minimalistic model}
Based on our experimental observations, we constructed a simple minimalistic model, which provides a semi-quantitative description of the dynamics of particle pairs. Our experiments have shown that hydrodynamic coupling and electrostatic interaction between particles are the key factors influencing the formation of stable colloidal pairs and strings. To induce the hydrodynamic coupling, the glass bottom provides the resistance force that slows down the motion of particles relative to the ambient flow. 

A full theoretical description of the motion of solid spheres close to a solid boundary in a semi-infinite fluid medium cannot be achieved within the limit of low-Re Stokes hydrodynamics \cite{Goldman67_1,Goldman67_2}. To fully address the problem, complex factors including the roughness of particles' surface \cite{Goldman67_1,Lin15}, the permeability and the stiffness of the layer of grafted PHSA polymer brushes \cite{Potanin95}, and even the cavitation of fluids underneath particles need to be considered \cite{Goldman67_1,Ashmore05}, which is certainly beyond the scope of our current work. Instead, we try to build a simple model, aiming for an order-of-magnitude estimate of our observations. Surprisingly, the minimalistic model shows a semi-quantitative agreement with our experiments. 

In our model, we reduce the complex effects of the bottom boundary on the motion of particles into one essential effect, i.e., it provides a resistance force, $f$, slowing down the motion of particles. The effect enables the hydrodynamic coupling between particles, crucial for the formation of colloidal strings. We further assume $f$ arises from the friction, instead of lubrication, between particles and the glass bottom and, therefore, is independent of the velocity of particles. Note that the density of particles matches that of the solvent. Hence, the normal force that results in the friction should arise from the electrostatic attraction between particles and the wall instead of the weight of the particles. Since the average surface charge of particles is approximately the same for different batches of particles (Sec. II), $f$ should also be independent of the size of particles.  The assumption is consistent with our experiments. At low flow rates, the drag on particles from the flow is below the threshold of static friction. Particles stick to the glass bottom and do not move (Fig.~\ref{equation2}a). At higher flow rates, particles slide on the glass bottom. The kinetic friction $f$ follows the Amonton-Coulomb law $f = \mu N$, where $\mu$ is the friction coefficient between PMMA spheres and the glass bottom and $N$ is the normal force induced by the electrostatic attraction between the spheres and the glass bottom. When salt is added into the solvent, the particle-wall interaction becomes repulsive, which removes the normal force. As a result, $f \approx 0$. Particles move passively with the flow (Fig.~\ref{figure6}b). The hydrodynamic coupling is zero and the string structures cannot be observed.

Under the above approximation, the force balance of the two particles in a particle pair can be easily written (Fig.~\ref{figure10} inset). Note that at low Reynolds number, the total force on each particle is zero. Therefore, for the leading particle (particle 1), we have
\begin{equation}
F_{h1} + F_e = f.
\label{equation4}
\end{equation}   
For the trailing particle (particle 2), we have 
\begin{equation}
F_{h2} = F_e + f.
\label{equation5}
\end{equation}
Here, $F_{hi}$ is the hydrodynamic force on particle $i$ and $F_e$ is the electrostatic repulsion between the two particles. $i=1$ or $2$. 

For the hydrodynamic forces, we have \cite{Brenner65}
\begin{eqnarray}
\frac{F_{h1}}{\eta K_1} & = &  \frac{(v_f - v_{p1}) - [K_2 (v_f-v_{p2})/4\pi r]}{1-(K_1K_2)/(4\pi r)^2}, \nonumber \\ 
\frac{F_{h2}}{\eta K_2} & = & \frac{(v_f - v_{p2}) - [K_1 (v_f-v_{p1})/4\pi r]}{1-(K_1K_2)/(4\pi r)^2},
\label{equation6}
\end{eqnarray}
where $r$ is the center-to-center distance between the two particles, $v_f$ is the velocity of the unperturbed ambient flow at the center of particles and $v_{pi}$ is the velocity of particle $i$. For a stable particle pair, $v_{p1} = v_{p2}$. The two particles move together with the same speed.  $K_i = 6\pi a_i$ is the coefficient of Stokes' drag of particle $i$. When $v_f = v_{pi}$, the hydrodynamic forces disappear. Furthermore, as $r \to \infty$, $F_{hi}$ reduces to $F_{hi} = \eta K_i(v_f-v_{pi})$, i.e., the Stokes law of isolated particles. The two particles decouple in this limit. 

For the electrostatic repulsion, the DLVO theory gives \cite{Russel89}
\begin{equation}
F_e = \frac{Z_1Z_2e^2}{\epsilon_r \epsilon_0}\frac{\exp(\kappa a_1)}{1+\kappa a_1}\frac{\exp(\kappa a_2)}{1+\kappa a_2}\frac{(1+\kappa r)}{r^2}\exp(-\kappa r),
\label{equation7}
\end{equation}
where $Z_i$ and $a_i$ are the number of charges and the radius of particle $i$, $e$ is the elementary charge, $\epsilon_r$ is the relative dielectric constant of the solvent, $\epsilon_0$ is the vacuum permittivity and $\kappa^{-1}$ is the screening length of the solvent. Since $Z_1Z_2>0$, $F_e$ is always repulsive.  

The force balance (Eqs.~\ref{equation4} and \ref{equation5}) leads to a simple relation,  
\begin{equation}
F_{h2} - F_{h1} = 2F_e.
\label{equation8}
\end{equation}   
Eq.~\ref{equation8} gives a zero internal force between the particles, $(F_{h2}-F_{h1}-2F_e)/2 = 0$. For two particles of the same size, $F_{h1} = F_{h2}$. In this case, Eq.~\ref{equation8} can be satisfied only by $F_e = 0$ when the distance between the two particles is significantly larger than the screening length $\kappa^{-1}$. With the two particles traveling at the same speed, the particle pair thus formed is metastable, susceptible to random perturbations. For stable particle pairs with nonzero $F_e$, polydispersity is necessary. Since $F_e>0$ and $F_{h1} < F_{h2}$, the leading particle should be smaller than the trailing particle in a particle pair, consistent with our observation (Fig.~\ref{figure9}). Eq.~\ref{equation8} determines the equilibrium position $r_0$ of particle pairs.   

\begin{figure}
\begin{center}
\includegraphics[width=3.1 in]{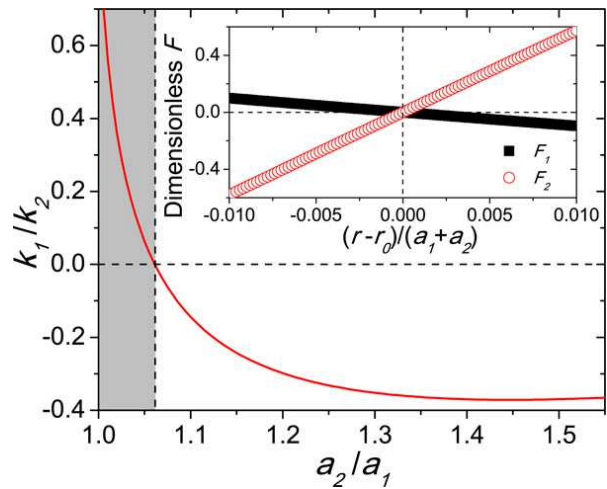}
\end{center}
\caption[Force on particles]{Effective spring-like forces on particle pairs from the model. The main plot shows the ratio of the effective spring constants, $k_1/k_2$, on two particles in particle pairs of different size ratios, $a_2/a_1$. The gray region indicates unstable particle pairs with $a_2/a_1 < 1.062$, where $k_1/k_2 > 0$. Inset shows the forces on the two particles in a particle pair with $a_2/a_1 = 1.11$ when the distance between the two particles $r$ is slightly away from their equilibrium position $r_0$. The forces are normalized by $k_BT/(a_1+a_2)$.} \label{figure11}
\end{figure}  

The prediction of $r_0$ from Eq.~\ref{equation8} is compared with experimental results in Fig.~\ref{figure8}. Since the leading particles are always small particles from the particle batch with average diameter $d=1.36$ $\mu$m, we fix $\langle a_1 \rangle = d/2 = 0.68$ $\mu$m in our calculation. In addition, we use $\langle Z_{1,2} \rangle = 157$ and the average velocity difference between particles and the ambient flow at $\langle v_f-v_p \rangle = 65$ $\mu$m/s, both of which are measured directly from experiments. The model qualitatively predicts the decreasing trend of $r_0$ with increasing $a_2/a_1$ (the black solid line in Fig.~\ref{figure8}). The quantitative discrepancy may arise from the crude approximation we have made on the effects of the wall and also from the fact the hydrodynamic forces in Eq.~\ref{equation6} is accurate only when the distance between particles are large $r_0 \gg (a_1+a_2)$ \cite{Brenner65}. Interestingly, when we plot the surface-to-surface distance, $\lbrack r_0-(a_1+a_2) \rbrack$, from the model, it quantitatively agrees with our experiments on $r_0$ (the black dashed line in Fig.~\ref{figure8}). The result indicates that the particle distance used in Eq.~\ref{equation6} may be replaced by an effective distance $r_e$ when $r_0$ is comparable with $(a_1+a_2)$. Considering the increasing contribution of the lubrication stress in the gap between particles to the hydrodynamic coupling, $r_e$ should be smaller than the center-to-center distance $r_0$ and close to the surface-to-surface distance between particles. Indeed, when we replace $r_0$ with $r_e = \lbrack r_0-(a_1+a_2) \rbrack$ in Eq.~\ref{equation6}, the result shows a quantitative agreement with our measurements (the red line in Fig.~\ref{figure8}).   

In addition to the equilibrium position $r_0$, the model also predicts that the approximately harmonic potential near $r_0$ is narrower and stiffer at larger $a_2/a_1$, leading to the decrease of the correlation time $\tau$ with $a_2/a_1$, consistent with our observation (Fig.~\ref{figure9} inset). Quantitatively, the force on the particle $i$ in a particle pair can be approximated as a spring-like force $F_{i} = k_{i}(r-r_0)$ when the distance between the two particles is slightly different from $r_0$, $|r-r_0|/r_0 \ll 1$ (Fig.~\ref{figure11} inset), where $k_{i}$ is the effective spring constant. $F_{i}$ is balanced by the Langevin-type random force from thermal fluctuations. The two effective springs on the particles in a particle pair are in series. Thus, the average spring constant of the effective force between the two particles can be simply estimated as $k = |k_1k_2|/(|k_1|+|k_2|)$. When we set the average particle ratio at $a_2/a_1 = 1.11$, the polydispersity of the particle batch used in Fig.~\ref{figure7}, the resulting spring-like force quantitatively matches our experiments (the black line in Fig.~\ref{figure7}c). The correlation time can also be estimated as $\tau \sim 1/\sqrt{k}$. A fitting using this model prediction again agrees well with our experiments at large $a_2/a_1$ (Fig.~\ref{figure9} inset). The predicted $\tau$, however, shows a sharper increase at small $a_2/a_1$. This discrepancy can be related to the limit of the stability of particle pairs at small $a_2/a_1$. 

To have a stable particle pair, the sign of $k_1$ and $k_2$ must be opposite, so that the forces acting on the particles are attractive when $r > r_0$ and repulsive when $r < r_0$ (Fig.~\ref{figure11} inset).  Based on the criterion, a particle pair becomes unstable or metastable when the size ratio of the two particles is smaller than 1.062 (Fig.~\ref{figure11}). The potential well becomes infinite soft and $\tau$ diverges at this lower bound of $a_2/a_1$. Due to the approximation taken in the model, the lower bound fails to explain the existence of particle pairs of small $a_2/a_1$ observed in our experiments, which may be metastable in nature. As a result, the predicted $\tau$ shows a sharper increase at small $a_2/a_1$ compared with experiments. Interestingly, the estimate lower bound quantitatively matches the size ratio of the most populous particle pairs found in our experiments at $a_2/a_1 = 1.064$ (Fig.~\ref{figure10}). 

Our simple minimalistic model is able to capture semi-quantitatively all the essential features of experimental observations. Although the slowdown of particle motions near the boundary is induced by the electrostatic attractions in our experiments, the model shows that any other attractive interactions such as those induced by commensurate polymer brushes attached to particles and walls \cite{Balazs06} can also lead to a similar mechanism for the assembly of colloidal strings. The non-equilibrium self-assembly principle explored here should apply in much broader contexts.        

\begin{figure}
\begin{center}
\includegraphics[width=3.3 in]{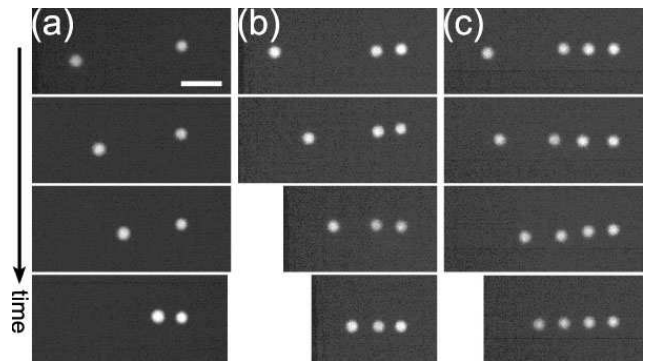}
\end{center}
\caption[Nucleation of long strings]{Nucleation and growth of 1D colloidal crystals. Fluid flows toward the right. (a)  The time series showing the formation of a particle pair (see also Supplementary Video 3). The times of the frames are 0 s, 0.53 s, 0.9 s and 1.91 s, respectively. (b) The time series showing the formation of a colloidal string of three particles (see also Supplementary Video 4). The times of the frames are 0 s, 1.85 s, 2.89 s and 5.42 s, respectively. (c) The time series showing the formation of a colloidal string of four particles (see also Supplementary Video 5). The times of the frames are 0 s, 0.78 s, 1.58 s and 2.46 s, respectively. The flow rate is 3.0 $\mu$L/min in the wide microfluidic channel. The scale bar is 7 $\mu$m.} \label{figure12}
\end{figure}         

\subsection{D. Kinetics and formation of colloidal strings}
The size difference between the leading and trailing particles also explains the kinetics of the formation of particle pairs. For an isolated particle, the particle velocity is given by $v_p = v_f - f/(6\pi\eta a)$ from the Stokes' law, where $f$ is the friction that is independent of the radius of the particle $a$. The larger size of the trailing particle results in a faster particle speed, which allows the large particle catch the small particle from behind and forms a stable particle pair (Fig.~\ref{figure12}a and Supplementary Video 3).   

\begin{figure}
\begin{center}
\includegraphics[width=3 in]{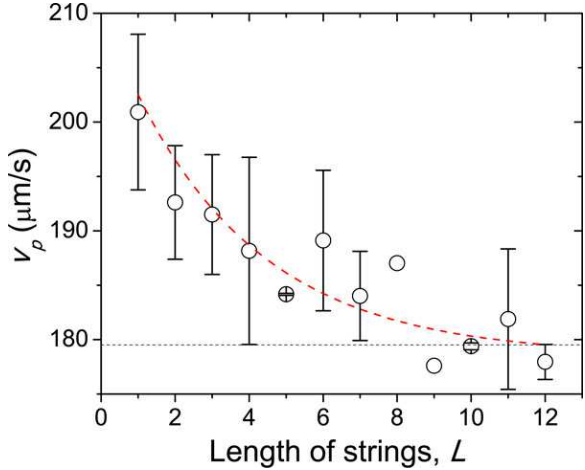}
\end{center}
\caption[Particle speeds]{The speed of particles in strings of different lengths, $L$, in terms of the number of particles. Flow rate is 3.0 $\mu$L/min. Particle speeds are measured at a fixed $y$ location in the narrow channel. The horizontal dashed line indicates the average speed of particles in strings with $L \geq 16$. The red line is a fit of an exponential function as a visual guide.} \label{figure13}
\end{figure}  

Particle pairs serve as the nuclei for the formation of longer colloidal strings at higher particle concentrations. Due to the hydrodynamic coupling, the speed of particles in colloidal strings decreases with the length of strings (Fig.~\ref{figure13}). For example, for a pair of equal-sized particles, the speed of particles is given by $v_p = v_f - c f/(6\pi\eta a)$ with $c = 1+3a/(2r) > 1$ (Eq.~\ref{equation6}), which is smaller than the speed of isolated particles. As a result, a third isolated particle can approach a particle pair from behind and form a colloidal string of three particles (Fig.~\ref{figure12}b and Supplementary Video 4). The same mechanism also applies for strings of four particles (Fig.~\ref{figure12}c and Supplementary Video 5). As the process repeats, a long colloidal string eventually forms. The dynamic assembly of long colloidal strings can be seen as a nucleation and growth process of 1D colloidal crystals in a flow-driven nonequilibrium system.                

\begin{figure}
\begin{center}
\includegraphics[width=3 in]{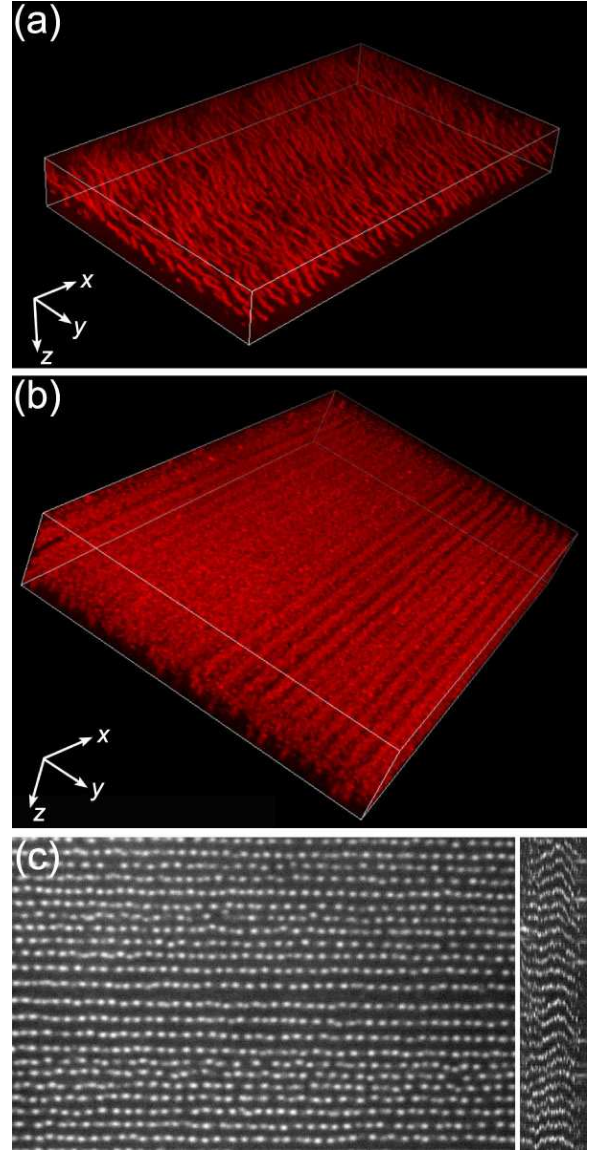}
\end{center}
\caption[Particle walls]{Formation of 2D colloidal walls in a transverse AC electric field. (a) A 3D view of colloidal chains formed in the AC field. The field of view is 120 $\times$ 75 $\times$ 14 $\mu$m$^3$ ($x \times y \times z$). The AC field is along the $z$ direction. (b) A 3D view of colloidal walls formed in the AC field under a fluid flow. The field of view is 120 $\times$ 75 $\times$ 16 $\mu$m$^3$ ($x \times y \times z$). The AC field is along the $z$ direction and the flow is along the $x$ direction. The walls move at 132 $\mu$m/s along the flow direction. (c) An orthogonal view of the colloidal walls near the center of the channel in the $x-y$ plane (left panel) and in the $z-y$ plane (right panel). The field of view of the left panel is 120 $\times$ 75 $\mu$m$^2$ and that of the right panel is 15 $\times$ 75 $\mu$m$^2$. The left and right edges of the right panel correspond to the bottom and top walls of the microfluidic channel made of ITO glasses.} \label{figure14}
\end{figure}   

\subsection{E. Construction of colloidal walls}
The hydrodynamic coupling that is responsible for the formation of 1D colloidal strings can also lead to the assembly of 2D colloidal walls in fluid flows when a transverse electric field is applied. To apply a transverse AC field across the channel along the $z$ direction, we construct a small microfluidic channel with an inner cross-section 100 $\times$ 15 $\mu$m$^2$ (W $\times$ H), where the top and bottom walls of the channel are made of ITO coated glasses. The small height of the channel was chosen here, so that we can apply a field with a large enough voltage gradient across the channel. We fixed the peak-to-peak amplitude of the field at 20 V (voltage gradient at 1333.33 V/mm) and the frequency at 50 kHz. 

Due to the field-induced dipolar interactions between particles, PMMA spheres suspended in CXB align into chains along the direction of the applied field (Fig.~\ref{figure14}a) \cite{Gast89,Parthasarathy96,Vutukuri12}. Note that we use pure CXB, instead CXB/decalin mixture, as our solvent in these experiments to enlarge the mismatch of dielectric constants between particles and the solvent, which increases the induced particle polarization. The formation of colloidal chains along the field direction results in the well-studied electrorheological effect \cite{Gast89,Parthasarathy96}. Without external fluid flows, the positions of the chains are completely random in the $x$-$y$ plane (Fig.~\ref{figure14}a). To induce the hydrodynamic coupling between different chains, a fluid flow is created along the $x$ direction by a syringe pump. Remarkably, under flow, the originally randomly-positioned chains align along with the flow direction and form 2D colloidal walls across the entire width of the channel in the $y$ direction (Figs.~\ref{figure14}b and c). The colloidal walls move along with the flow but with significantly slower speeds (Supplementary Video 6). The spacing between the walls are controlled by the particle concentration.          

\section{IV. Conclusions}
We have conducted an experimental study on the dynamic self-assembly of charged colloidal particles in microfluidic flows. We showed that due to the balance of hydrodynamic coupling and electrostatic repulsion, colloidal particles can self-assemble into 1D colloidal strings close to the boundary of microfluidic channels. The strings are stable against strong perturbations in the system. Such a 1D ordered structure can be further extended into 2D colloidal walls when a transverse electric field is imposed. We systematically investigated both the dynamics and the kinetics of the formation of the 1D colloidal strings and constructed a simple minimalistic model that provides a semi-quantitative explanation of the self-assembly process. Our experiments demonstrate a new way for ordering micron-sized objects in low-Re flows, potentially useful for various diagnostic microfluidic devices. One possible strategy to exploit this unique non-equilibrium self-assembly process is to construct microfluidic channels of narrow heights, which increases the surface-to-volume ratio of the channels and facilitates the boundary-induced self-assembly of colloidal strings. 

A future experiment is planned to investigate the kinetics of the formation of 2D colloidal walls. Moreover, the complex effects of the boundary on the hydrodynamic coupling and electrostatic interaction need to be further studied, which will provide a better quantitative description of the dynamic self-assembly process.            

\section{Acknowledgments}
We thank Pranav Agrawal, Kevin Dorfman, Woo Jin Hyun, Satish Kumar, Truong Pham, Yi Peng and Seunghwan Shin for the help with experiments and fruitful discussions. This research was supported by the University of Minnesota Industrial Partnership for Research in Interfacial and Materials Engineering (IPRIME). Y.A. acknowledges the financial support from the Toray Industries. L. G. was partially supported by Conicyt PAI/Postdoctorado Becas Chile 74150032. The research was partially supported by the David $\&$ Lucile Packard Foundation. Portions of this work were performed in the University of Minnesota NanoFabrication Center, which receives partial support from the NSF through the NNIN.

\end{document}